\documentclass[twocolumn,amssymb,nobibnotes,aps,showpacs]{revtex4}
\usepackage{amsthm,amsfonts,amssymb,amsmath,dsfont}
\usepackage[latin1]{inputenc}
\usepackage[pdftex]{graphicx,graphics,color}
\usepackage[english]{babel}
\usepackage{indentfirst}
\usepackage{hyperref}

\begin{document}
\renewcommand\arraystretch{1.5}
\newcommand{\nn}{\nonumber}
\newcommand{\Tr}{\mathrm{Tr}}
\newcommand{\Ln}{\mathrm{Ln}}
\newcommand{\bra}{\langle}
\newcommand{\ket}{\rangle}
\newcommand{\del}{\partial}
\newcommand{\vt}{\vec}
\newcommand{\dg}{^{\dag}}
\renewcommand{\Re}{\mathrm{Re}}
\renewcommand{\Im}{\mathrm{Im}}
\renewcommand{\mod}{\mathrm{\,mod\,}}

\renewcommand{\topfraction}{0.85}
\renewcommand{\textfraction}{0.1}

\title{Phase Transition, Entanglement and Squeezing in a Triple-Well Condensate}

\author{T. F. Viscondi}
\affiliation{Institute of Physics ``Gleb Wataghin'',
University of Campinas, P. O. Box 6165, 13083-970, Campinas, SP, Brazil}
\author{K. Furuya}
\affiliation{Institute of Physics ``Gleb Wataghin'',
University of Campinas, P. O. Box 6165, 13083-970, Campinas, SP, Brazil}
\author{M. C. de Oliveira}
\affiliation{Institute of Physics ``Gleb Wataghin'',
University of Campinas, P. O. Box 6165, 13083-970, Campinas, SP, Brazil}

\begin{abstract}
We provide an in-depth characterization of a three modes Bose-Einstein condensate trapped in a symmetric circular triple well
potential. We analyze how a subsystem independent measure of entanglement, the purity related to the $su(3)$ algebra, scales for
increasing number of atoms and signals correctly the quantum phase transition between two dynamical regimes in a specific
arrangement. Moreover, this measure, which is intrinsically related to particle entanglement, also depicts if some squeezing is
occurring when we consider the system's ground state.
\end{abstract}


\maketitle
Entanglement has played an important role for the understanding of quantum many body aspects \cite{amico08} that
traditionally belonged to statistical mechanics and quantum field theory. Several investigations in quantum critical models at
$T=0$ have shown that complex entangled ground state contains all the important correlations that give rise to different phases
known to exist in several systems \cite{qpt}. Thus, it is a fact that entanglement study in many body { systems} allows a deeper
characterization of the ground state of the system undergoing a quantum phase transition (QPT), particularly its order.
Characterization of a QPT via pairwise and collective multipartite entanglement has been given in a very conclusive way in Refs.
\cite{wu04,oliveira06}, but are dependent on the specific partition employed, i.e., they are subsystem dependent. Another way of
investigating subsystem independent entanglement in many-particle systems is through the generalized purity associated to the
pertinent algebra \cite{barnum03}. Beyond being a measure for the quality of semiclassical approximation \cite{viscondi09}, this
measure is also related to squeezing of moments of the generators of the pertinent algebra. Recently both { spin} squeezing and
entanglement have been demonstrated for a $^{87}$Rb condensate trapped in double and multiple wells of an optical lattice
\cite{obt08}, and subsystem independent entanglement theoretically investigated in Ref. \cite{viscondi09} for a double-well
trapped condensate. In fact although those results were developed independently they are profoundly complementary since they
relate QPT, entanglement and squeezing, for a system which is a particular realization of the Lipkin-Meshkov-Glick model
\cite{LMG,vidal04}. The interplay between entanglement and squeezing has been investigated previously in many instances
\cite{spinsqueezing}, and now seems to play an important role in QPT involving many bosons as well.

In this Letter we investigate in detail a BEC of attractively interacting neutral atoms trapped in a symmetric triple well
potential in a three mode approximation and show that the ground state of the model undergoes a QPT. A time dependent variational
principle using $SU(3)$ coherent state allows for a system of semiclassical equations that  enables one to find the fixed points
of the model and to investigate how the lowest energy fixed points change as the collision parameters of the model are varied.
Since the lowest energy state in this system corresponds to a twin condensate fixed point, where effectively the system behaves
as if composed of two wells (although coherences between the twin modes are still present) we can show that the $SU(3)$ reduces
to $SU(2)$ while in this regime. The non-linear components of the Hamiltonian provided by the interactions of our model lead to
spin squeezing for the ground state as we vary the scattering parameter, which on its turn is associated to variance squeezing of
some of the generators of the algebra, and shows non-trivial phenomena such as particle entanglement related to QPT and to
distinct dynamical regimes. This special arrangement is promising for experimental investigation with existent trapping
technology.

The model of a Bose-Einstein condensate trapped in a triple well potential has already been considered in literature in many
distinct configurations \cite{nemoto00,penna04}. Considering that the coupling between the potential wells is sufficiently weak,
we shall use the usual local modes approximation \cite{milburn97} associated to the states $|u_{j}\ket$, $j=1,2,3$, representing
the ground states of harmonic approximations around each minimum of the trap. Considering this approximation to be valid even
when the total number of trapped bosons $N$ is large, the overlap $\varepsilon\equiv \bra u_{i} |u_{j}\ket$, ($i\neq j$) must be
very small. Keeping terms up to $O(\varepsilon^{\frac{3}{2}})$, and using the fact that the total number of trapped particles
$N$ is conserved, we can write the Hamiltonian:

\begin{equation}
H= \Omega^{'} \sum\limits_{i\neq
j}a\dg_{i}a_{j}+\kappa\sum\limits_{i}a^{\dag 2}_{i}a^{2}_{i}
-2\Lambda{\sum\limits_{i\neq j\neq k}}a\dg_{i}a_{i}a\dg_{j}a_{k};
\label{eq1}
\end{equation}

\noindent where $\Omega^{'}=\Omega+2\Lambda(N-1)$, and the operator $a_{j} (a^{\dagger}_j)$, $j=1,2,3$, annihilates (creates) a
boson in the state $|u_{j}\ket$. In the last equation we also defined the parameters of the system: $\Omega$ is the tunneling
rate, $\Omega^{'}$ is the effective tunneling rate, $\kappa$ is the self-collision parameter and
$\Lambda=\kappa\varepsilon^{\frac{3} {2}}\ll\kappa$ is the cross-collision rate, which is proportional to the interaction
frequency between bosons in different sites. Although $\Lambda$ is a lower order parameter of the model, we see that its presence
in the effective tunneling rate $\Omega'\equiv\Omega+2\Lambda(N-1)$ is relevant for $N\gg1$.

Since the Hamiltonian \eqref{eq1} preserves $N$, we can take advantage of the homomorphism between the commutation relations of
$SU(3)$ group generators and bilinear combinations of creation and annihilation operators in the three local modes, making use of
an extension of Schwinger's pseudo-spin method \cite{schwinger65}:

\begin{align}
Q_{1}&\equiv\frac{1}{2}(a\dg_{1}a_{1}-a\dg_{2}a_{2}),&
Q_{2}&\equiv\frac{1}{3}(a\dg_{1}a_{1}+a\dg_{2}a_{2}-2a\dg_{3}a_{3});\nonumber\\
J_{k}&\equiv i(a\dg_{k}a_{j}-a\dg_{3}a_{j}),&
P_{k}&\equiv a\dg_{k}a_{j}+a\dg_{2}a_{j};
\label{eq2}
\end{align}

\noindent for $k =1,2,3$ and $j=(k+1)\mod3+1$. Using these eight generators we can rewrite the Hamiltonian:

\begin{eqnarray}
H &=&\left(\Omega'-2\Lambda\frac{N}{3}\right)
(P_{1}+P_{2}+P_{3})+\frac{\kappa}{2}(4Q_{1}^{2}+3Q_{2}^{2})\nonumber\\
&&+\Lambda\left[2Q_{1}(P_{1}-P_{3})+Q_{2}(2P_{2}-P_{1}-P_{3})\right],
\label{eq3}
\end{eqnarray}

showing that $SU(3)$ is the pertinent dynamical group. Note that the tunneling term is linear in the generators, while the
collision terms are quadratic.

A semiclassical dynamics of the condensate can be developed by using the time dependent variational principle \cite{saraceno81} {
with  coherent states of the group $SU(3)$ \cite{perelomov86} as test functions,} allowing us to express the semiclassical
Hamiltonian as $\mathcal{H}(\vt{w}^{*},\vt{w})\equiv\bra N;\vt{w}|H|N;\vt{w}\ket$, with the coherent states expanded on the Fock
space basis for the three local modes as

\begin{equation}
|N;\vt{w}\ket= \displaystyle{C\sum\limits_{n_{1}+n_{2}+n_{3}=N}}\sqrt{\frac{N!}{n_{1}!n_{2}!n_{3}!}}
w_{1}^{n_{1}}w_{2}^{n_{2}}|n_{1},n_{2},n_{3}\ket,
\label{eq4}
\end{equation}

\noindent where $C=1/(|w_{1}|^{2}+|w_{2}|^{2}+1)^{\frac{N}{2}}$. The vector $\vt{w}=(w_{1},w_{2})\in\mathds{C}^{2}$ parametrizes
the non-linear subspace of Hilbert space consisting only of coherent states and also represents a point in phase space associated
with the classical analogue system \cite{zhang89}. The { semiclassical} equations of motion  are then brought to the Hamilton's
canonical form by the transformation of variables $w_{j}=\sqrt{I_{j}/(N-I_{1}-I_{2})}e^{-i\phi_{j}};\quad j=1,2.$ While $I_{j}$
is the average occupancy in the $j$-th local mode, $\phi_{j}$ represents the difference in phase between the macroscopic
condensates located in the $j$-th and the third wells. The semiclassical approximation becomes exact if the Hamiltonian is linear
in the generators of the group or in the classical limit of the system, which coincides with the macroscopic limit
$N\rightarrow\infty$ \cite{yaffe82}. The phase space position and the stability of the equilibrium points of the equations of
motion depend { only} on the model parameters  $\chi\equiv\frac{\kappa(N-1)}{\Omega}$ and $\mu\equiv\frac{\Lambda(N-1)}{\Omega}$
(See Ref. \cite{thiago09b}).

As previously noticed in Ref. \cite{penna04} there are fixed point solutions corresponding to configurations where two localized
condensates are in phase and have the same occupational average, which are described by the three equivalent conditions
$w_{1}=w_{2}$, $w_{1}=1$ or $w_{2}=1$, known as restrictions of \textit{twin condensates}. Many of the fixed points of the
semiclassical dynamics are contained in this sub-regime under the additional condition of $\phi_{1},\phi_{2}=0,\pi$, which
implies that $\vt{w}\in\mathds{R}^{2}$. Without loss of generality, we choose only the condition $w=w_{1}=w_{2}\in\mathds{R}$, so
that the $SU(3)$ coherent states reduce to coherent states of $SU(2)$, which thus brings many features observed for a two mode
condensate, such as the Rabi Oscillation (RO) of population and the macroscopic self-trapping (MST) of population
\cite{thiago09}. The phase space associated with the integrable sub-regime of twin condensates is isomorphic to $S^{2}$, a space
that parametrizes the set of { $SU(2)$ coherent states $|J=\frac{N}{2}, \tau=\sqrt{2}w=
\tan{\frac{\theta}{2}}e^{-i\phi}\rangle$}. An interesting quantity for the dynamical analysis in the twin condensates sub-regime
is the population balance $I_{z}\equiv-\cos\theta=\frac{4I_{1}-N}{N}$. \vspace*{-1cm}

\begin{widetext}
\begin{center}
\begin{figure}[htbp]
\includegraphics[height=4.8cm]{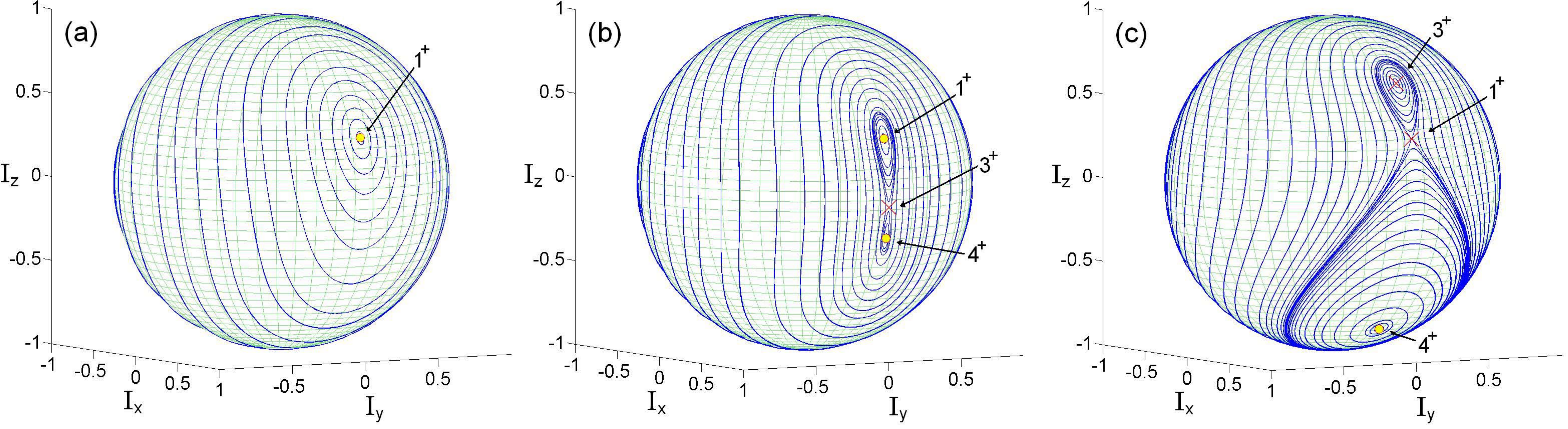}
\begin{minipage}{\textwidth}
\caption{\footnotesize (color online) Semiclassical trajectories in the sub-regime of twin condensates on the unit sphere for
various initial conditions. For the three spheres we use $N=30$, $\Omega=-1$ and $\mu=0$, while the values of the reduced
self-collision parameter are: $(a)$ $\chi=1.5$, $(b)$ $\chi=1.98$ and $(c)$ $\chi=3$. The yellow circles (red crosses) indicate
points of stable (unstable) equilibrium for the full dynamics. The critical parameter of bifurcation in the absence of
cross-collisions is $\chi_{+}(\mu=0)\approx1.97$.} \label{fig1}
\end{minipage}\vspace{-1cm}
\end{figure}
\end{center}
\end{widetext}

The point $w_{1}=w_{2}=1$ is a solution of the fixed point equations, independent of the values of $\chi$ and $\mu$, which we
call $1^{+}$. Considering always $\mu\ll\chi$,  we have another solution if $\chi<\chi_{+}(\mu)$, which we call $2^{+}$; whereas
if  $\chi>\chi_{+}(\mu)$ two further solutions of the fixed point equations exists. In other words, a \textit{saddle-node}
bifurcation with critical parameter $\chi_{+}(\mu)$ is responsible for the appearance of the fixed points $3^{+}$ and $4^{+}$.

Fig. \ref{fig1}.a shows the semiclassical dynamics of twin condensates in the absence of the fixed points $3^{+}$ and $4^{+}$. We
see that all trajectories are around the fixed point $1^{+}$ and therefore vary around the value $I_{z}=\frac{1}{3}$, which
represents same average occupancy in the three local modes. This is exactly the RO dynamical regime, where the system does not
show any preferential mode occupation. In figure \ref{fig1}.b, where $\chi$ exceeds the critical value of bifurcation, we observe
the emergence of new types of orbits in the system, especially around the new stable equilibrium point $4^{+}$. These new
trajectories oscillate around negative values of $I_{z}$ and show a suppression of tunneling between the twin condensates and the
solitary one, resulting in the preferential occupation of the third mode. Such a dynamical effect is known as \textit{macroscopic
self-trapping} (MST), similar to the case of two modes \cite{milburn97,smerzi97}. As the value of $\chi$ increases, the region of
phase space occupied by orbits associated with the MST regime also grows larger, as we see in figure \ref{fig1}.c.

A quantitative error in the semiclassical approximation is originated when we force the state to evolve preserving minimal
uncertainty on the space phase, {\sl i.e.}, obligating the system to remain as a coherent state. If the quantum dynamics of the
system leads to an increase of the state uncertainty, we cannot expect a high quantitative accuracy in the semiclassical
treatment. Including the situation where strong squeezing or spreading over the phase space occurs, leading thus to a large
deviation of the coherent state. An effective measure of the quality of the semiclassical approximation is the generalized purity
associated with the $su(3)$ algebra \cite{barnum03}, defined by

\begin{equation}
\begin{array}{rcl}
\mathcal{P}_{su(3)}(|\psi\ket) &=& \frac{9}{N^{2}}
\Bigl( \frac{\bra\psi|Q_{1}|\psi\ket^{2}}{3}+
\frac{\bra\psi|Q_{2}|\psi\ket^{2}}{4} \\
&&+\sum\limits_{j=1}^{3}\frac{\bra\psi|P_{j}|\psi\ket^{2}}{12}
+\sum\limits_{k=1}^{3}\frac{\bra\psi|J_{k}|\psi\ket^{2}}{12} \Bigr).
\end{array}
\label{eq10}
\end{equation}

This measure is derived from the total uncertainty of the algebra \cite{delbourgo77} and has maximum value $\mathcal{P}_{su
(3)}(|\psi\ket)=1$ only if $|\psi\ket$ is a coherent state of $SU(3)$, corresponding to the state of minimum total uncertainty.
In contrast, the purity presents a minimum value $\mathcal{P}_{su(3)}(|\psi\ket)=0$ for states of largest uncertainty. Therefore,
$\mathcal{P}_{su(3)}$ can be used to measure the ``distance'' of a particular state to the subspace formed only by the coherent
states. That is, the larger  the uncertainty of a state during its evolution, the lower  is $\mathcal{P}_{su (3)}$ and less
accurate is the semiclassical approximation using only coherent states \cite{viscondi09}.

In addition the generalized purity is also a genuine measure of separability in systems of many identical particles. Under a
basis transformation of the single-particle Hilbert space, every coherent state of the type \eqref{eq4} can be rewritten as a
separable state in each boson:

\begin{equation}
\begin{split}
|N;\vt{w}\ket
=\frac{1}{\sqrt{N!}}\left[\frac{w_{1}a_{1}\dg+w_{2}a_{2}\dg+a_{3}\dg}
{(|w_{1}|^{2}+|w_{2}|^{2}+1)^{\frac{1}{2}}}\right]^{N}|0\ket\\
=\frac{(a_{\vt{w}}\dg)^{N}}{\sqrt{N!}}|0\ket\;=\;|N_{\vt{w}}\ket
\;=\;\bigotimes\limits_{i=1}^{N}|\psi_{\vt{w}}^{(i)}\ket;
\label{eq11}
\end{split}
\end{equation}

\noindent where $|\psi_{\vt{w}}\ket\equiv\frac{w_{1}|u_{1}\ket+w_{2}|u_{2}\ket+|u_{3}\ket}
{(|w_{1}|^{2}+|w_{2}|^{2}+1)^{\frac{1}{2}}}$ is an arbitrary state of the single-particle Hilbert space \footnote{Except by a
choice of global phase and normalization.} with associated annihilation operator $a_{\vt{w}}$. Due to the symmetrization
principle, we see that \eqref{eq11} is the only possibility of a completely separable state in each particle, considering systems
of many identical bosons. Therefore, the states with $\mathcal{P}_{su(3)}(|\psi\ket)=1$ represent the separable states of $N$
bosons, while the decrease of purity from this maximum value indicates the entanglement among the particles of the system. In
this sense when the state of the system is less localized in the phase space, $\mathcal{P}_{su(3)}(|\psi\ket)<1$,  it is more
entangled. Several dynamical processes can be responsible for the loss of purity of an initially coherent state. One of such
processes is the squeezing of the state, and it is related to regular dynamical regimes of the semiclassical model
\cite{thiago09}. In the semiclassical approximation we also observe chaos \cite{thiago09b} in the three mode model, accompanied
by other processes of purity loss.

The generalized purity can also be written as the trace of a squared reduced density operator, however this reduction is done
with respect to the algebra. Considering $\rho=|\psi\ket\bra\psi|$, we have
$\mathcal{P}_{su(3)}(|\psi\ket)=K\Tr(\rho_{su(3)}^{2})$, where $K$ is a normalization factor dependent on $N$, $\rho_{su
(3)}=\sum\limits_{j}\Tr(\rho A_{j})A_{j}$ is the density operator reduced in $su(3)$ and $\{A_{j}\}$ is a basis of $su(3)$
orthonormalized in relation to the trace in the Fock space of $N$ particles. Therefore, the entanglement can be described as the
decoherence of the system in its dynamical algebra. Summarizing, pure (mixed) states in the algebra are equivalent to separable
(entangled) states in each boson. The semiclassical approximation is accurate only if the state remains separable in each boson,
when we can completely describe the system by only one trajectory in the phase space that extremizes an action functional. Then,
the entanglement is the fundamental quantum feature responsible for breaking-off the classicality of the system, and it signals
QPT as we discuss now.

\begin{figure}[htbp]
\centering
\includegraphics[height=4.5cm]{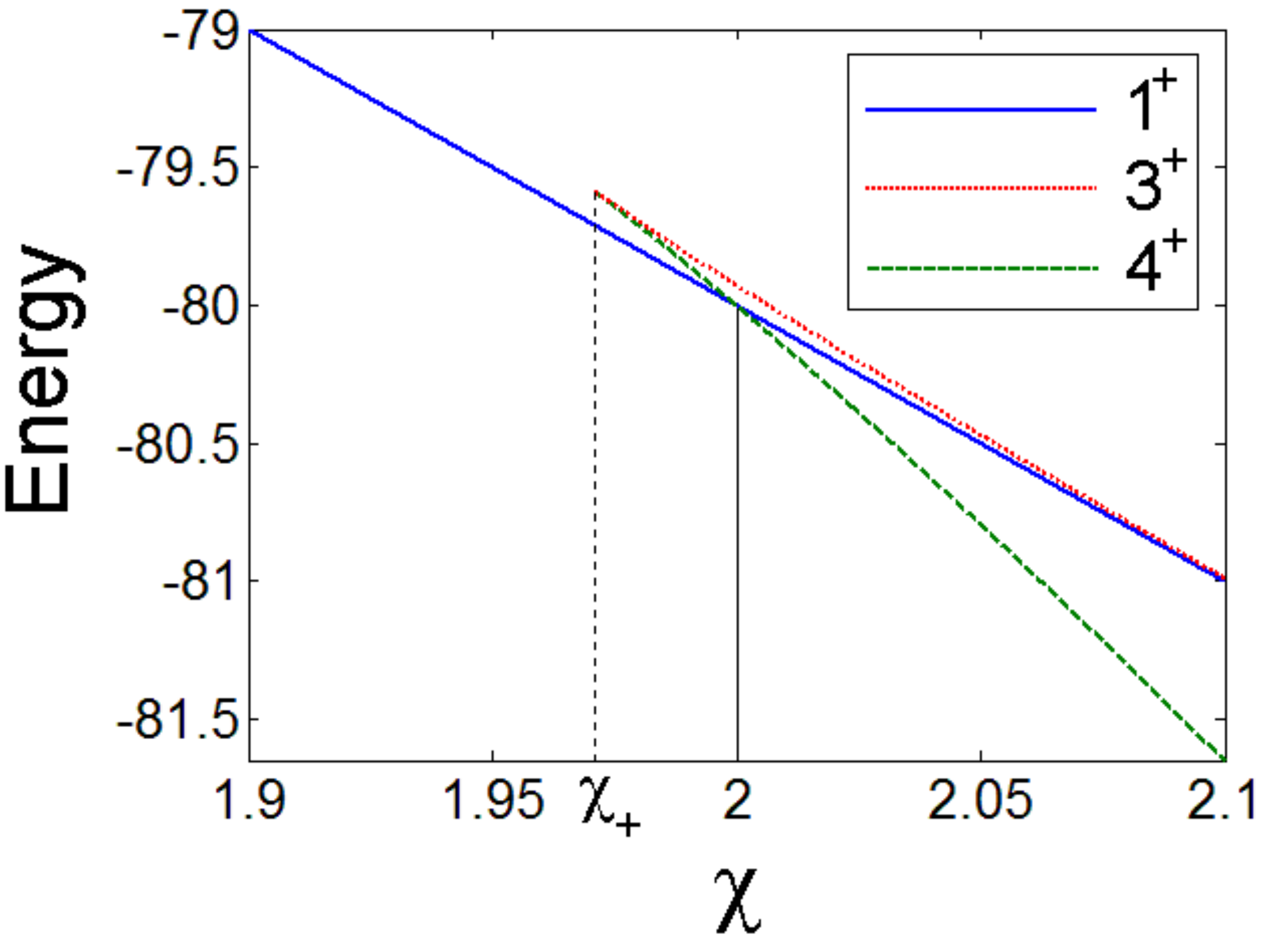}
\caption{\footnotesize (color online) $\mathcal{H}$ calculated at its extremes as a function of the self-collision parameter
$\kappa$ showing the fixed points for the twin-condensate regime for $N=30$, $\Omega=-1$ and $\mu=0$. }
\label{fig2}
\end{figure}

QPT results from non-analyticities in the ground state energy as a function of a real parameter of the Hamiltonian, characterized
only in the macroscopic limit $N\rightarrow\infty$ and at zero temperature. However, as stressed earlier, the macroscopic limit
of the model is equivalent to its classical limit, so that the minimum of the semiclassical energy per particle  $\frac{{\cal
H}}{N}$, which is independent of $N$, can be used exactly to study the QPT. The extremes of $\mathcal{H}$ are the equilibrium
points of the model. In Fig. \ref{fig2} we show $\mathcal{H}$ calculated in each of its extremes as a function of $\chi$, with
$N=30$, $\Omega=-1$ and $\mu=0$. The occurrence of a \textit{first order} QPT is evident at the critical value $\chi_{c}=2$,
since there is a discontinuity of the first derivative of the ground state energy, resulting from the energy level crossing of
the fixed points $1^{+}$ and $4^{+}$.

\begin{figure}[htbp]
\centering
\includegraphics[height=11.5cm]{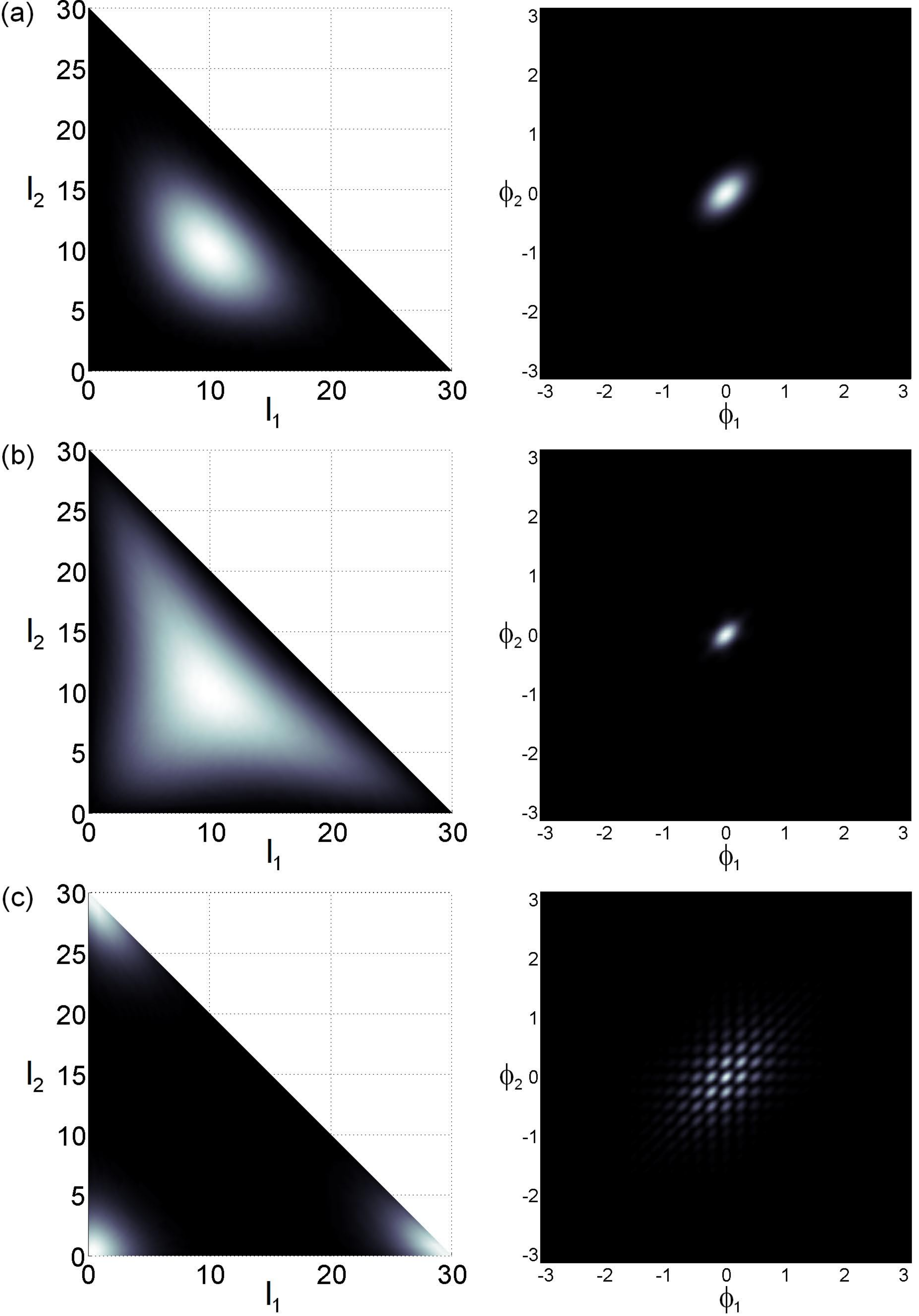}
\caption{\footnotesize Occupational Husimi representation $Q_{I}(I_{1},I_{2})$ (above) and phase distribution
$\Phi(\phi_{1},\phi_{2})$ (below) of the ground state for $N=30$, $\Omega=-1$, $\mu=0$ and self-collision parameter value: $(a)$
$\chi=0$, $(b)$ $\chi=2$ and $(c)$ $\chi=3$. Remark the squeezing in phase while the distribution in number broadens. The
fragmentation of the phase in (c) characterizes the QPT.} \label{fig3}
\end{figure}

As already shown, the fixed point $1^{+}$ is the center of the orbits related to the RO dynamical regime, characterized by no
preferential occupation of the local modes. The three equivalent equilibrium points $4^{+}$, one in each sub-regime of twin
condensates, are responsible for the emergence of semiclassical trajectories associated with the MST regime in the solitary mode.
Therefore, we expect the ground state to display characteristics of the RO (MST) regime for $\chi<\chi_{c}$ ($\chi>\chi_{c}$).
Furthermore, the behavior of the low energy equilibrium points suggests a non-degenerate (triply degenerate) ground level when
the self-collision parameter is lower (higher) than $\chi_{c}$ in the limit $N\rightarrow\infty$.

Phase space representation of projections of the ground state into subspaces demonstrate signals of the quantum phase transition
for finite $N$, as shown in figure \ref{fig3}. The population Husimi function
$Q_{I}(I_{1},I_{2})\equiv\int\limits_{0}^{2\pi}\int\limits_{0}^{2\pi}\frac{d\phi_{1}d\phi_{2}} {(2\pi)^{2}}|\bra
N;\vt{w}|\psi\ket|^{2}$ is a quasi-probability distribution of the average occupations in the three local modes, while the
function $\Phi(\phi_{1},\phi_{2})\equiv|\bra\phi_{1},\phi_{2}|\psi\ket|^{2}$, with
$|\phi_{1},\phi_{2}\ket\equiv\sum\limits_{n_{1}+n_{2}+n_{3}=N} e^{-in_{1}\phi_{1}-in_{2}\phi_{2}}|n_{1},n_{2},n_{3}\ket$,
represents the probability distribution of collective phase differences \cite{barnett86} between the local condensates.
Considering $\chi=\mu=0$ in Fig. \ref{fig3}.a, the exact ground state is the coherent state $|N;w_{1}=w_{2}=1\ket$, centered at
the equilibrium point $1^{+}$, showing the non-preferential occupation of the local modes. For increasing value of $\chi$ in Fig.
\ref{fig3}.b, we observe the state expansion in the population subspace, accompanied by squeezing in the conjugate subspace,
related thus to phase squeezing. Note that the coherent states are the most localized states in the phase space, but not
necessarily in a subspace. When $\chi=\chi_{c}$ there is no abrupt change in the representations of the ground state, since such
changes should happen in a continuous way for finite $N$. When the self-collision parameter takes values slightly larger than
$\chi_{c}$ we notice a profound change in the ground state, shown in Fig. \ref{fig3}.c as a \textit{trifurcation} of the
occupational distribution, each of the components is more squeezed than the original coherent state. This is a accompanied by the
emergence of an {\sl interference pattern} in the conjugate subspace. This \textit{fragmentation} of $\Phi$ represents the
superposition in states of preferential occupancy in each local mode, also visible in the behavior of $Q_{I}$. Notice that the
ground state representations in phase space display a smooth transition from RO to MST regime, in contrast to the abrupt
transition shown in the macroscopic limit.

\begin{figure}[htbp]
\centering
\includegraphics[height=6cm]{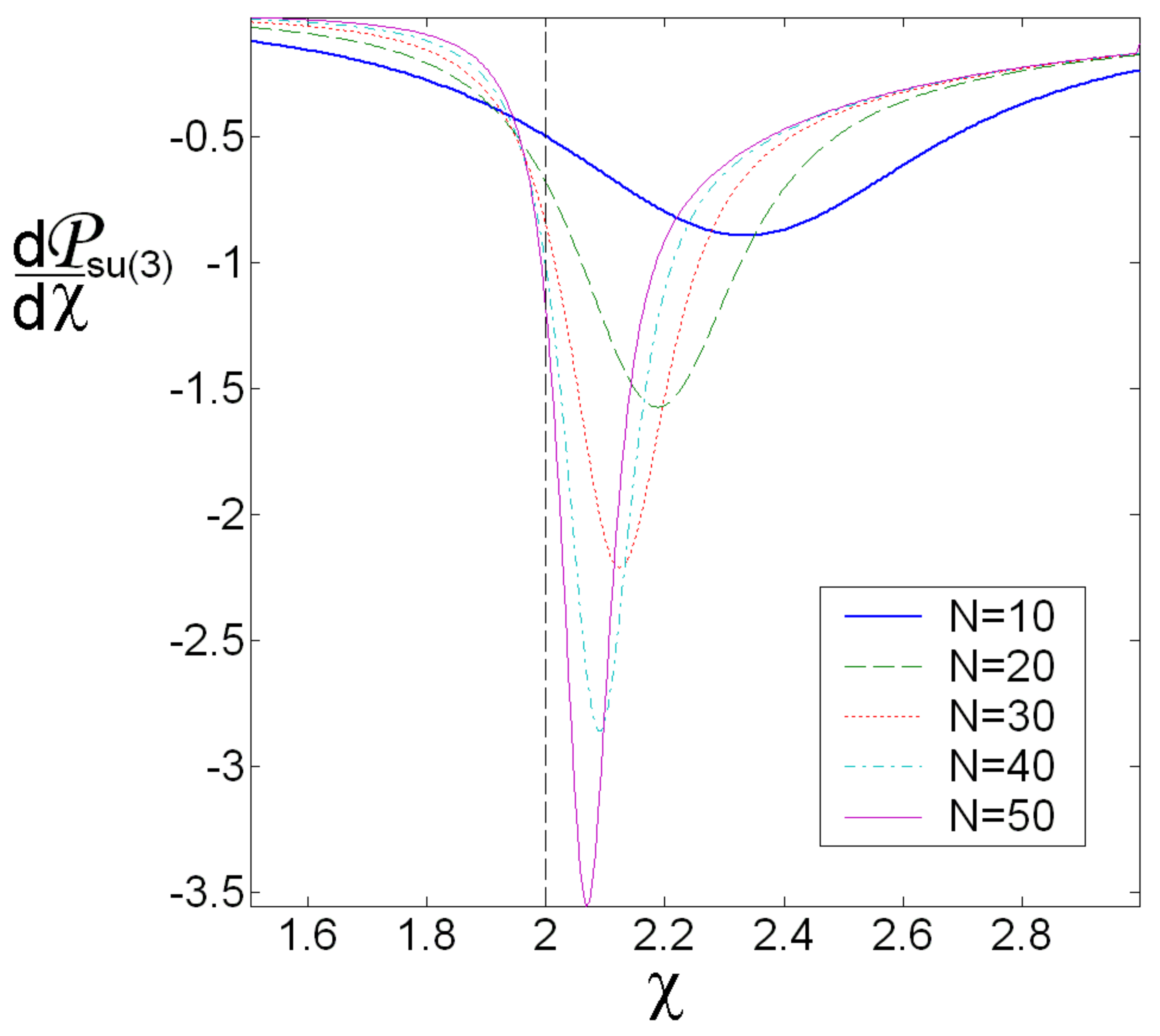}
\caption{\footnotesize (color online) We show $\frac{d\mathcal{P}_{su(3)}}{d\chi}$ calculated in the ground state as a function
of $\chi$, for $\Omega=-1 $, $\mu=0$ and several values of $N$. $\chi_{c}$ is indicated by the dashed vertical line.}
\label{fig4}
\end{figure}

Now we return to the discussion of particle entanglement, as given by the generalized purity, and the signaling of the QPT
\cite{wu04}. In figure \ref{fig4} we show $\frac{d\mathcal{P}_{su(3)}}{d\chi}$ calculated at the ground state as a function of
$\chi$, considering several values of $N$. In the absence of bosonic collisions the ground state is a coherent state and
consequently its purity value is equal to one. For $\chi<\chi_{c}$, the region related to the expansion-compression of the ground
state representations, we see that $\mathcal{P}_{su (3)}$ decreases slowly with increasing $\chi$. However, for values of $\chi$
slightly above $\chi_{c}$ we observe a rapid decay of $\mathcal{P}_{su (3)}$ caused by the ground state fragmentation associated
with the tunneling suppression. The loss of purity is faster for larger number of trapped bosons, suggesting a scaling property
between $\frac{d\mathcal{P}_{su(3)}}{d\chi}$ and $N$. We define the \textit{scalable quantum critical parameter}
$\chi_{c}^{q}(N)$ as the self-collision parameter value that minimizes $\frac{d\mathcal{P}_{su(3)}}{d\chi} $ for a specific
number of particles.  We observe that the minimum value of $\frac{d\mathcal{P}_{su(3)}}{d\chi}$ becomes more pronounced for
increasing $N$, while $\chi_{c}^{q}(N)$ moves to the left, toward $\chi_{c}$. The values of $\chi_{c}^{q}(N)$ obtained from
figure \ref{fig4} suggest a linear relationship between $\ln[\chi_{c}^{q}(N)-\chi_{c}]$ and $\ln(N)$. A linear interpolation of
the data provides the following power-law: $\chi_{c}^{q}(N)-\chi_{c}=e^{1.2\pm0.2}N^{-0.99\pm 0.05}$. Therefore, the convergence
law of $\chi_{c}^{q}(N)$ to $\chi_{c}=2$ as $N\rightarrow\infty$ is characterized. Also $\mathcal{P}_{su(3)}$ signals the QPT
correctly, considering the scaling behavior for finite $N$ \footnote{In the classical-macroscopic limit $\mathcal{P}_{su(3)}$ is
not well defined as a signal of QPT because the ground level is triply degenerate after the transition and there is no unique
choice of state to calculate the purity, giving different results for different linear combinations of degenerate ground states.
This difficulty does not exist for finite $N$, as the ground state is
unique.}.

Although the critical bifurcation parameter $\chi_{+}$ is distinct from the critical transition parameter $\chi_{c}$, the
emergence of new fixed points in the semiclassical dynamics shows a clear relation with the occurrence of the QPT. This same
characteristic has been observed in the two local modes model \cite{hines03}, but with some important differences. The two-mode
model presents a \textit{second order} QPT, accompanied by a \textit{pitchfork} bifurcation with critical parameter equal to the
critical  parameter $\chi_{+}^{TM}=\chi_{c}^{TM}=\frac{1}{2}$ \cite{thiago09}, in the absence of cross-collisions.

\begin{figure}[tbhp]
\centering
\includegraphics[height=5cm]{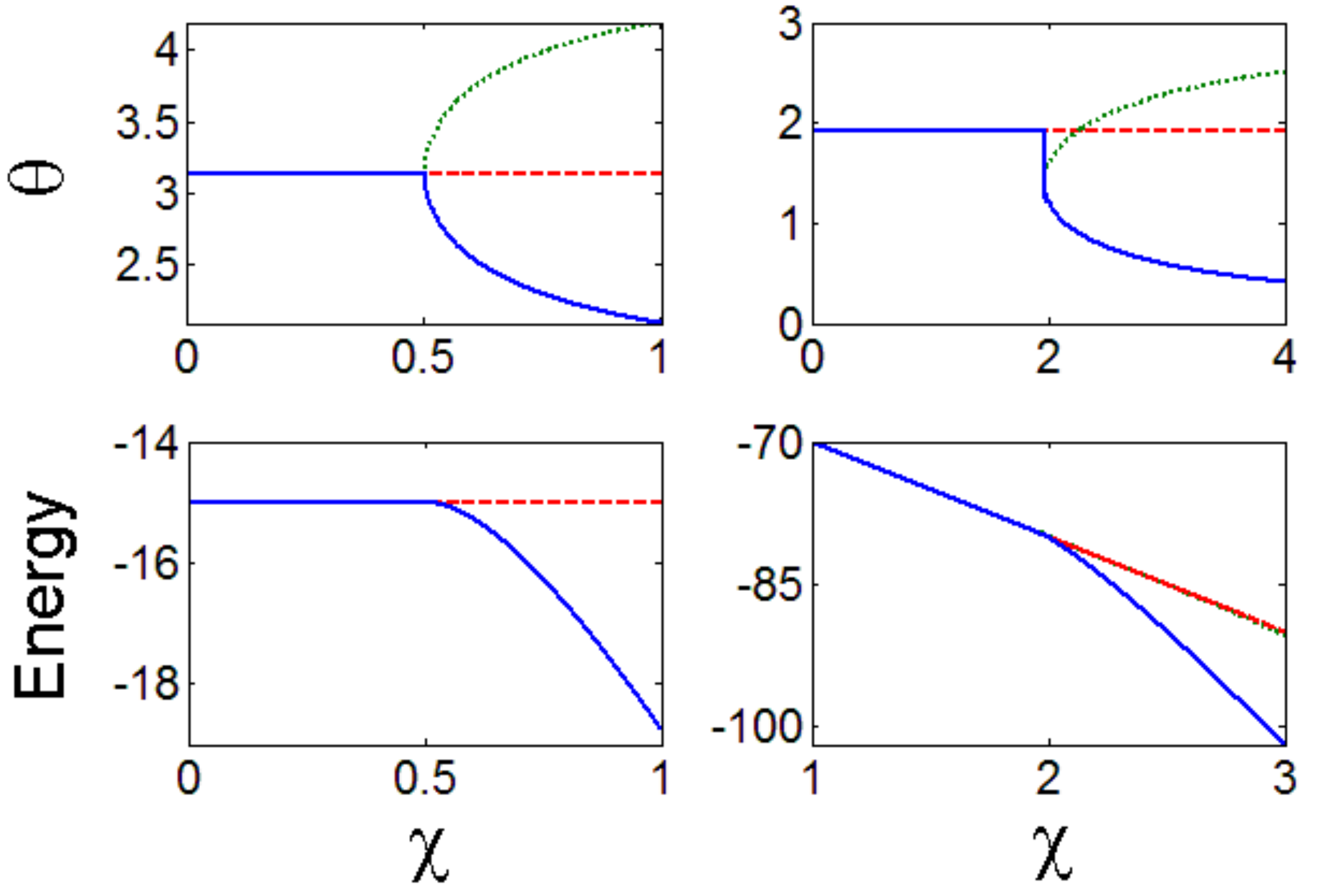}
\caption{\footnotesize (color online) On the left the solid (blue) lines show the curves of $\theta_{min}$ (above) and
$\mathcal{H}(\theta_{min})$ (below) considering the two local modes model. On the right we show the same curves for the
three-mode model. Remark that differently from the two modes, there is an discontinuity in $\theta$ for the three modes.}
\label{fig5}
\end{figure}

However, in both models the equilibrium point that minimizes the semiclassical energy $\mathcal{H}$ can be described by a single
angle $\theta$ on $ S^{2}$, { considered as a function of $\chi$}. For example, the restriction $w_{1}=w_{2}\in\mathds{R}$
determines the unidimensional manifold of the phase space that contains the minimum energy fixed point. Therefore, we generically
consider $\mathcal{H}$ restricted to only one dimension in the analysis of the QPT:
$\mathcal{H}(\theta;\chi)=\mathcal{H}_{0}(\theta)+\chi\mathcal{H}_{1}(\theta)$; where $ \mathcal{H}_{0}$ ($ \mathcal {H}_{1}$) is
the portion of the semiclassical Hamiltonian obtained from the linear (quadratic) terms of $H$ in the generators of the dynamical
group. The position of the semiclassical ground state $\theta_{min}\equiv\theta_{min}(\chi)$, which satisfies the necessary
conditions $\left.\frac{\del\mathcal{H}}{\del\theta}\right|_{\theta=\theta_{min}}=0$ and
$\left.\frac{\del^{2}\mathcal{H}}{\del\theta^{2}}\right|_{\theta=\theta_{min}}>0$, and the minimum energy curve
$\mathcal{H}_{min}(\chi)\equiv\mathcal{H}(\theta_{min}(\chi),\chi)$ are shown in figure \ref{fig5} for the two and three-mode
models \footnote{The minimum energy level is doubly (triply) degenerated after the transition in the two (three) local modes
model, but we can restrict the phase space so that the choice of $\theta_{min}(\chi)$ is unique. There is no loss of generality,
because the minima are equivalent.}. The function $\theta_{min}(\chi)$ possesses discontinuous first derivative at
$\chi_{c}^{TM}$ for the two-mode model, while for the three-mode model we observe a discontinuity in the curve of
$\theta_{min}(\chi)$ itself at the critical value $\chi_{c}$. The different types of bifurcation are responsible for the distinct
behavior of the curve $\theta_{min}(\chi) $ in both models, resulting also in different orders of transition. The first
derivative of $\mathcal{H}_{min}$ with respect to the parameter $\chi$ is given by:

\begin{equation}
\frac{d\mathcal{H}_{min}}{d\chi}=\left.\frac{\del\mathcal{H}}{\del
\theta}\right|_{\theta=\theta_{min}}\frac{d\theta_{min}}{d\chi}
+\left.\frac{\del\mathcal{H}}{\del\chi}\right|_{\theta=\theta_{min}}=\mathcal{H}_{1}(\theta_{min})
\label{eq14}
\end{equation}

Therefore, the continuity of $\theta_{min}$ implies that $\frac{d\mathcal{H}_{min}}{d\chi}$ is also continuous, since
$\mathcal{H}_{1}(\theta_{min})$ is a continuous function of $\theta_{min}$. In general, the reciprocal is also true, because
$\mathcal{H}_{1}(\theta_{min})$ is discontinuous in the case of $\theta_{min}$ also discontinuous, resulting in a first order
QPT. The discontinuity in the derivative of $\theta_{min}$ can be identified only in the second derivative of
$\mathcal{H}_{min}$:

\begin{equation}
\begin{aligned}
\frac{d^{2}\mathcal{H}_{min}}{d\chi^{2}}&=
\left.\frac{d\mathcal{H}_{1}}{d\theta}\right|_{\theta=\theta_{min}}\frac{d\theta_{min}}{d\chi}
\end{aligned}
\label{eq15}
\end{equation}

So, in general, the discontinuity of $\frac{d\theta_{min}}{d\chi}$ implies discontinuity only for the second derivative of
$\mathcal{H}_{min}$, resulting in a second-order QPT.

In conclusion we have described the dynamics of a  BEC trapped in a symmetric triple well potential in a three mode approximation
and show in details the way in that the ground state of the model undergoes a QPT. A time dependent variational principle using
$SU(3)$ coherent state allows for a system of semiclassical equations that  enables one to find the fixed points of the model and
to investigate how the lowest energy fixed points change as the collision parameters of the model are varied. Thus particle
entanglement signals correctly the presence of QPT as one crosses distinct dynamical regimes. We have shown that the ground state
of the model is related to the subdynamics of the twin condensate, a regime whose dynamics falls into the $SU(2)$ group. We also
show that the increase in the entanglement of the ground state is associated to phase compression (Rabi oscillation) and to
number compression (self-trapping). In this way we relate the loss of coherence in the fundamental state to ``spin squeezing''.
Remark that the squeezing is not the only effect that leads to the decrease of the generalized purity. Whenever there is a
departure from a coherent state the purity decreases, but this behavior unambiguously describe particle entanglement in the
system. The more entangled are the particles the smaller is the purity associated to the pertinent algebra. We remark that this
three symmetric well model allows for a better characterization of QPT than the case for double well, providing not only the
dynamical transitions but signaling the presence of squeezing and thus could arouse experimental investigation with existent
trapping technology for this phenomenon. This amount of squeezing and entanglement, if controllable, could provide a new scenario
for testing quantum physics fundamental questions.

We acknowledge Markus Oberthaler for many clarifying discussions and valuable comments on our results. This work was partially
supported by FAPESP and CNPq through the Brazilian National Institute of Science and Technology on Quantum Information.

\end{document}